\documentstyle[12pt,epsfig]{article}
\def \app{D_{\pi \pi}}

\def \bea{\begin{eqnarray}}
\def \beq{\begin{equation}}
\def \bo{B^0}

\def \cn{Collaboration}
\def \cpp{C_{\pi \pi}}
\def \eea{\end{eqnarray}}
\def \eeq{\end{equation}}
\def \ite{{\it et al.}}

\def \lpp{\lambda_{\pi \pi}}
\def \ob{\overline{B}^0}

\def \rpp{R_{\pi \pi}}

\def \spp{S_{\pi \pi}}
\begin{document}

\begin{flushright}
TECHNION-PH-2002-12\\
EFI 02-62 \\
hep-ph/0202170 \\
February 2002 \\
\end{flushright}

\renewcommand{\thesection}{\Roman{section}}
\renewcommand{\thetable}{\Roman{table}}
\centerline{\bf STRONG AND WEAK PHASES FROM TIME-DEPENDENT}
\centerline{\bf MEASUREMENTS OF $B \to \pi \pi$}
\medskip
\centerline{Michael Gronau}
\centerline{\it Physics Department, Technion -- Israel Institute of Technology}
\centerline{\it 32000 Haifa, Israel}
\medskip
\centerline{Jonathan L. Rosner}
\centerline{\it Enrico Fermi Institute and Department of Physics}
\centerline{\it University of Chicago, Chicago, Illinois 60637}
\bigskip

\begin{quote}

Time-dependence in $B^0(t) \to \pi^+ \pi^-$ and $\ob(t) \to \pi^+ \pi^-$ is
utilized to obtain a maximal set of information on strong and weak phases.  One
can thereby check theoretical predictions of a small strong phase $\delta$
between penguin and tree amplitudes.  A discrete ambiguity between $\delta
\simeq 0$ and $\delta \simeq \pi$ may be resolved by comparing the observed
charge-averaged branching ratio predicted for the tree amplitude alone, using
measurements of $B \to \pi l \nu$ and factorization, or by direct comparison 
of Cabibbo-Kobayashi-Maskawa (CKM) matrix parameters with those determined by
other means.  It is found that with 150 fb$^{-1}$ from BaBar and Belle, this
ambiguity will be resolvable if no direct CP violation is found.  In the
presence of direct CP violation, the discrete ambiguity between $\delta$ and
$\pi - \delta$ becomes less important, vanishing altogether as $|\delta| \to
\pi/2$.  The role of measurements involving the lifetime difference between
neutral $B$ eigenstates is mentioned briefly.
\end{quote}

\leftline{\qquad PACS codes:  12.15.Hh, 12.15.Ji, 13.25.Hw, 14.40.Nd}

\section{Introduction}

The observation of CP violation in decays of $B$ mesons to $J/\psi$ and
neutral kaons \cite{betaBa,bCSBe} has inaugurated a new era in the study of
matter-antimatter asymmetries.  Previously, such asymmetries had been
manifested only in the decays of neutral kaons and in the baryon asymmetry of
the Universe.  CP violation in $B$ and neutral kaon decays is described
satisfactorily in terms of phases in the Cabibbo-Kobayashi-Maskawa (CKM)
matrix, but the baryon asymmetry of the Universe apparently requires sources of
CP violation beyond the CKM phases.  There is thus great interest in testing
the self-consistency of the CKM description through a variety of processes.

One key test of the CKM picture involves the decays $B^0 \to \pi^+
\pi^-$.  The time-dependence of $B^0|_{\rm initial} \to
\pi^+ \pi^-$ and $\ob|_{\rm initial} \to \pi^+ \pi^-$ involves 
quantities $\spp$ and $\cpp$ which are, respectively,
coefficients of terms involving $\sin \Delta m t$ and $\cos \Delta m t$,
and which depend in different ways on strong and weak phases.  The
BaBar Collaboration reported the first measurement of these quantities
\cite{CSBa1}, recently updated to $\spp = -0.05 \pm 0.37 \pm 0.07$ and $\cpp =
-0.02 \pm 0.29 \pm 0.07$ \cite{CSBa2}.  The Belle Collaboration reports $\spp =
-1.21^{+0.38+0.16}_{-0.27-0.13}$ and $\cpp= -0.94^{+0.31}_{-0.25} \pm 0.09$
\cite{bCSBe}, using BaBar's sign convention for $\cpp$.  The averages are
$\spp = -0.66 \pm 0.26$ and $\cpp = -0.49 \pm 0.21$.

Both model-independent considerations \cite{GR01,LRpipi} and explicit
calculations in QCD-improved factorization \cite{BBNS} indicate that
a crude measurement of $\spp$ around zero implies a significant
restriction on CKM parameters if
the strong phase difference $\delta$ between two amplitudes contributing
to $B^0 \to \pi^+ \pi^-$ is small ($\delta \simeq 10^\circ$ in
\cite{BBNS}; see, however, \cite{KLS}.) The quantity $\cpp$
provides information on $\delta$ if the phase and $\cpp$ are both near zero,
but a discrete ambiguity allows the phase to be near $\pi$ instead.

In the present paper we re-examine the decays $B^0 \to \pi^+ \pi^-$
to extract the maximum amount of information directly
from data rather than relying on theoretical calculations of strong phases.
We find that if $\sin \delta$ is small one can resolve a discrete
ambiguity between $\delta \simeq 0$ and $\delta \simeq \pi$ by comparing
the measured branching ratio of $B^0 \to \pi^+ \pi^-$ (averaged over
$B^0$ and $\ob$) with that predicted in the absence of the penguin
amplitude.  The latter can be obtained using 
information on the semileptonic process $B \to \pi l \nu$
assuming factorization for color-favored processes, which appears
to hold well under general circumstances \cite{LRfact}.

We find that with data foreseen within the next two years
it should be possible to reduce theoretical and
experimental errors to the level that a clear-cut choice can be made
between the theoretically-favored prediction of small $\delta$ and the
possibility of $\delta \simeq \pi$, assuming that the parameter $\cpp$
describing direct CP violation in $B^0 \to \pi^+ \pi^-$ remains
consistent with zero.  If $\cpp \sim \sin \delta$ is found to be
non-zero, direct CP violation will have been demonstrated in $B$ decays,
a significant achievement in itself.  The sign of $\cpp$ will then determine
the sign of $\delta$.  While the discrete ambiguity 
between $\delta$ and $\pi - \delta$ then becomes harder to resolve, its
effect on CKM parameters becomes less important.

We recall notation for $B^0 \to \pi^+ \pi^-$ decays in Sec.\ II.  The
dependence of $\spp$ and $\cpp$ on weak and strong phases is exhibited
in Sec.\ III.  It is seen that when $|\cpp|$ is maximal, there is little
effect of any discrete ambiguity, since the strong phase $\delta$ is
close to $\pm \pi/2$, while when $\cpp \simeq 0$ the discrete ambiguity
between $\delta \simeq 0$ and $\delta \simeq \pi$ results in very
different inferred weak phases.  The use of the flavor-averaged $\bo \to
\pi^+ \pi^-$ branching ratio to resolve this ambiguity is discussed in
Sec.\ IV, while the CKM parameter restrictions implied by the observed
$\spp$ range are compared in Sec.\ V for $\delta = 0$ and $\delta =
\pi$.

One more observable, which we call $\app$, obeys $\spp^2 + \cpp^2 + \app^2 =
1$, so its magnitude is fixed by $\spp$ and $\cpp$, but its sign provides new
information.  In principle, it is measurable in the presence of a detectable
width difference between neutral $B$ meson mass eigenstates, as is shown in
Sec.\ VI.  However, we find that the sign of $\app$ is always negative for
the allowed range of CKM parameters, and does not help to resolve the
discrete ambiguity.  A positive value of $\app$ would signify new physics.
We conclude in Sec.\ VII.

\section{Notation}

We use the same notation as in Ref.\ \cite{GR01}, to which the reader is
referred for details.  We define $T$ to be a color-favored tree amplitude 
in $B^0 \to \pi^+\pi^-$
and $P$ to be a penguin amplitude \cite{GHLR}.  Using standard definitions of
weak phases (see, e.g., \cite{PDG}) $\alpha = \phi_2$, $\beta = \phi_1$,
and $\gamma = \phi_3$, the decay amplitudes to $\pi^+ \pi^-$ for $B^0$ and
$\ob$ are
$$
A(\bo \to \pi^+ \pi^-) = -(|T|e^{i \delta_T} e^{i \gamma} +
 |P| e^{i \delta_P})~~~,
$$
\beq \label{eqn:Bpipi}
A(\ob \to \pi^+ \pi^-) = -(|T|e^{i \delta_T} e^{- i \gamma} +
 |P| e^{i \delta_P})~~~,
\eeq
where $\delta_T$ and $\delta_P$ are strong phases of the tree and penguin
amplitudes, and $\delta \equiv \delta_P - \delta_T$.  Our convention
will be to take $-\pi \le \delta \le \pi$.  

The coefficients of $\sin \Delta m_d t$ and $\cos \Delta m_d t$ measured in
time-dependent CP asymmetries of $\pi^+ \pi^-$ states produced in asymmetric
$e^+ e^-$ collisions at the $\Upsilon(4S)$ are \cite{Gr}
\beq \label{eqn:CSpipi}
\spp \equiv \frac{2 {\rm Im}(\lpp)}{1 + |\lpp| ^2}~~,~~~
\cpp \equiv \frac{1 - |\lpp|^2}{1 + |\lpp|^2}~~~,
\eeq
where
\beq
\lpp \equiv e^{-2i \beta} \frac{A(\ob \to \pi^+ \pi^-)}
{A(B^0 \to \pi^+ \pi^-)}~~~.
\eeq
In addition we may define the quantity
\beq \label{eqn:Apipi}
\app \equiv \frac{2 {\rm Re}(\lpp)}{1 + |\lpp|^2}~~~,
\eeq
for which it is easily seen that
\beq
\spp^2 + \cpp^2 + \app^2 = 1~~,~~~{\rm implying}~~~
\spp^2 + \cpp^2 \le 1~~.
\eeq
The significance of $\app$ will be discussed in Sec.\ VI.

When $\delta=0$ or $\pi$ the quantity $\lpp$ becomes a pure phase:
\beq
\lpp = e^{2 i \alpha_{\rm eff}}~~,~~~
\alpha_{\rm eff} = \alpha + \Delta \alpha~~~,
\eeq
\beq
\Delta \alpha = \left\{ \begin{array}{c} \arctan \frac{|P/T| \sin
\gamma}{1 + |P/T| \cos \gamma}~~~(\delta = 0) \cr
- \arctan\frac{|P/T| \sin\gamma}{1 - |P/T| \cos \gamma}~~~(\delta =
\pi)~~~. \end{array} \right.
\eeq
In such cases $\spp = \sin(2 \alpha_{\rm eff})$, $\app = \cos(2
\alpha_{\rm eff})$.

The expressions (\ref{eqn:Bpipi}) employ the phase convention in which
top quarks are integrated out in the short-distance effective
Hamiltonian and the unitarity relation $V^*_{ub} V_{ud} + V^*_{cb}
V_{cd} = - V^*_{tb} V_{td}$ is used, with the $V^*_{ub} V_{ud}$ piece of
the penguin operator included in the tree amplitude \cite{GR01}.  Using
these expressions and substituting $\alpha = \pi - \beta - \gamma$, we then
may write
\beq
\lpp = e^{2 i \alpha} \left( \frac{1 + |P/T| e^{i \delta}
e^{i \gamma}}{1 + |P/T| e^{i \delta} e^{-i \gamma}} \right)~~.
\eeq
The consequences of assuming $\delta$ small, as predicted in Ref.\
\cite{BBNS}, were explored in Refs.\ \cite{GR01,LRpipi}.  In the former, it was
shown that even an earlier crude measurement \cite{CSBa1} of $\spp$, taken at
$1\sigma$, drastically reduced the allowed CKM parameter space.  In the latter,
where a slightly different convention for penguin amplitudes was used,
it was shown how to use $\spp$ and $\cpp$ to determine weak and strong
phases.

One needs a value of $|P/T|$ to apply these expressions to data.  In
Refs.\ \cite{GR01} and \cite{LRpipi} $|P|$ was estimated using
experimental data on $B^+ \to K^0 \pi^+$ (a process dominated by the
penguin amplitude aside from small annihilation contributions) and
flavor SU(3) including SU(3) symmetry breaking, while $|T|$ was estimated
using factorization and
data on $B \to \pi l \nu$.  We shall use the result of Ref.\ \cite{GR01},
$|P/T| = 0.276 \pm 0.064$.  Ref.\ \cite{BBNS} found $0.285 \pm 0.076$, which
included an estimate of annihilation, and Ref.\ \cite{LRpipi} obtained
$0.26 \pm 0.08$,  based on a different phase convention for the penguin
amplitude, without including SU(3) breaking effects.  The individual amplitudes
of Ref.\ \cite{GR01}, in a convention in which their square gives a
$B^0$ branching ratio in units of $10^{-6}$, are $|T| = 2.7 \pm 0.6$
and $|P| = 0.74 \pm 0.05$.  We shall make use of them in Sec.\ IV.

It is most convenient to express $\spp$, $\cpp$, and $\app$ in terms of
$\alpha$, $\beta$, and $\delta$, using $\gamma = \pi - \alpha - \beta$,
since when $P = 0$ one has $\spp = \sin 2 \alpha$ and 
$\app = \cos 2 \alpha$.  The value
of $\beta$ is fairly well known as a result of the recent measurements
by BaBar \cite{betaBa} and Belle \cite{bCSBe}:  $\sin 2 \beta = 0.78
\pm 0.08$, $\beta = (26 \pm 4)^\circ$.  Defining
\beq
\overline{{\cal B}}(B^0 \to \pi^+ \pi^-) \equiv
[{\cal B}(B^0 \to \pi^+ \pi^-) + {\cal B}(\ob  \to \pi^+ \pi^-)]/2~~~,
\eeq
\beq\label{rpp}
\rpp \equiv \frac{\overline{{\cal B}}(B^0 \to \pi^+ \pi^-)}
{\overline{{\cal B}}(B^0 \to \pi^+ \pi^-)|_{\rm tree}} = 1 - 2|P/T| \cos \delta
\cos(\alpha + \beta) + |P/T|^2~~~,
\eeq
explicit expressions for $\spp,~\cpp$, and $\app$ are then
\beq
\spp = [\sin 2 \alpha + 2 |P/T| \sin(\beta - \alpha) \cos \delta
- |P/T|^2 \sin 2 \beta]/\rpp~~
\eeq
\beq
\cpp = [2 |P/T| \sin (\alpha + \beta) \sin \delta]/\rpp~~,
\eeq
\beq
\app = [\cos 2 \alpha - 2 |P/T| \cos(\beta - \alpha) \cos \delta
+ |P/T|^2 \cos 2 \beta]/\rpp~~.
\eeq
The quantity $\rpp$ itself will be used in Sec.\ IV to resolve a
discrete ambiguity, while the usefulness of the sign of $\app$ will be
described in Sec.\ VI.

Note that $\cpp$ is odd in $\delta$ while $\spp$ and $\app$ are even in
$\delta$.  Within the present CKM framework one has 
$0 < \alpha + \beta < \pi$, implying $\sin (\alpha + \beta) > 0$, so that
a measurement of non-zero $\cpp$ will specify the sign
of $\delta$ (predicted in some theoretical schemes \cite{BBNS}).

We shall concentrate for the most part on a range of CKM parameters
allowed by fits to weak decays, disregarding the possibility of
new physics effects.  Aside from the constraints associated
with $\spp$, it was found in Ref.\ \cite{Beneke} (quoting \cite{Hocker}
and \cite{Ciu}; see also \cite{GR01}) that $\sin 2 \alpha = -0.24 \pm
0.72$, implying $\alpha = (97^{+30}_{-21})^\circ$, which we shall take
as the ``standard-model'' range.

One could regard the three equations for $\rpp$, $\spp$, and $\cpp$ as
specifying the three unknowns $|P/T|$, $\delta$, and $\alpha$ (given the
rather good information on $\beta$).  In what follows we shall, rather,
use the present constraints on $|P/T|$ mentioned above, first
concentrating on what can be learned from $\spp$ and $\cpp$ alone and
then using the information on $\rpp$ both as a consistency check and to
resolve discrete ambiguities.  The information provided by the sign of
$\app$ will be treated separately.

\begin{figure}
\centerline{\epsfysize = 4.7 in \epsffile{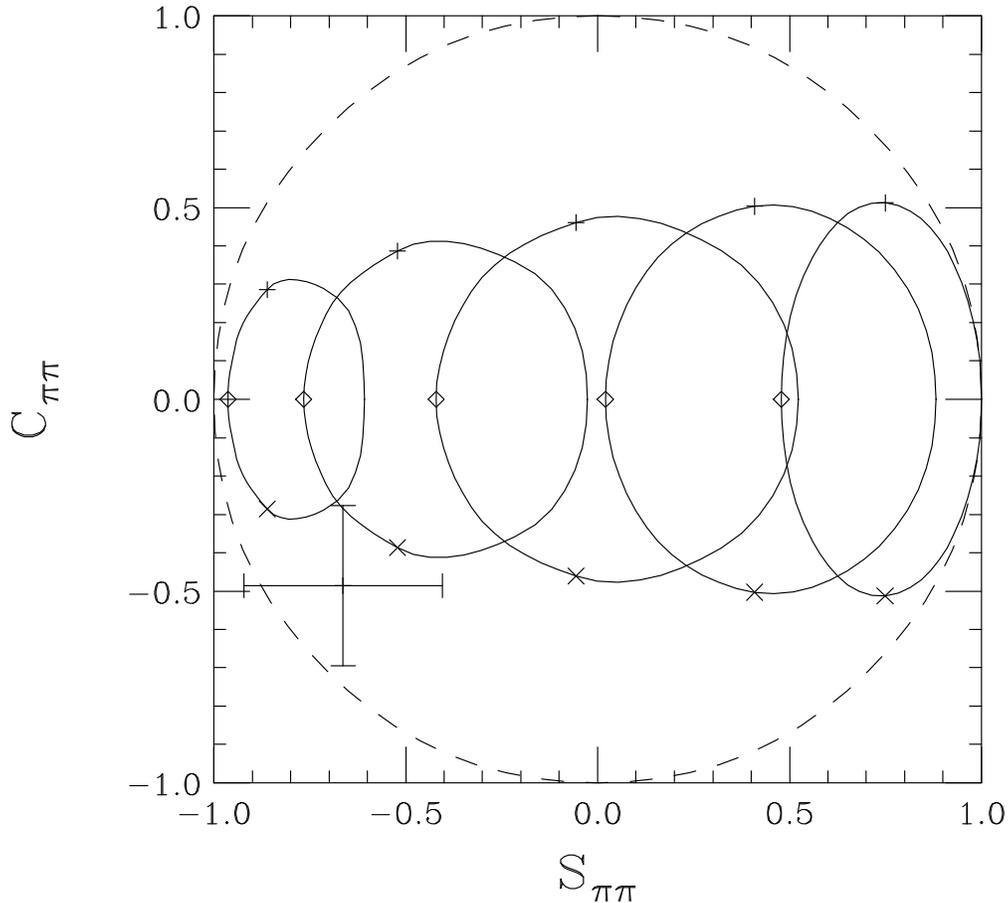}}
\caption{Values of $\spp$ and $\cpp$ for representative values of
$\alpha$ lying roughly in the physical region.  Closed curves
correspond, from right to left, to $\alpha = 60^\circ$, $75^\circ$,
$90^\circ$, $105^\circ$, and $120^\circ$.  Plotted points on curves
correspond to $\delta = 90^\circ$ ($+$ signs), 0 (diamonds),
and $-90^\circ$ (crosses).  The dashed circle denotes the bound
$\spp^2 + \cpp^2 \le 1$.  The plotted point with large errors
corresponds to the average of the measurements \cite{bCSBe,CSBa2} of $\spp$ and
$\cpp$.  The central values $\beta = 26^\circ$, $|P/T| = 0.28$ have been taken.
\label{fig:cs00}}
\end{figure}
 
\section{Dependence of $\spp$ and $\cpp$ on $\alpha$ and $\delta$}

We display in Fig.\ \ref{fig:cs00} the values of $\spp$ and $\cpp$ for $\alpha$
roughly in the physical region, with $-\pi \le \delta \le \pi$.  For any fixed
$\alpha$, the locus of such points is a closed curve with the points $\delta =
0$ and $\delta = \pm \pi$ corresponding to $\cpp = 0$ and with $\cpp(-\delta) =
- \cpp(\delta)$.  A large negative value of $\spp$, as seems to be indicated by
the Belle measurement \cite{bCSBe}, favors large values of $\alpha$. Negative
values of $\cpp$ imply a negative $\delta$.  The sum of squares of $\spp$ and
$\cpp$ is always bounded by 1, and one can show that for any value of $\delta$
and $\alpha + \beta$ one has $|\cpp| \le 2|P/T|/(1 + |P/T|^2)$.  For a {\it
given} value of $\alpha + \beta$ the bound is stronger:
\beq
|\cpp| \le \frac{2|P/T| |\sin(\alpha + \beta)|}{\sqrt{(1 + |P/T|^2)^2
- 4|P/T|^2 \cos^2(\alpha+\beta)}}~~~.
\eeq

\begin{figure}
\centerline{\epsfysize = 5 in \epsffile{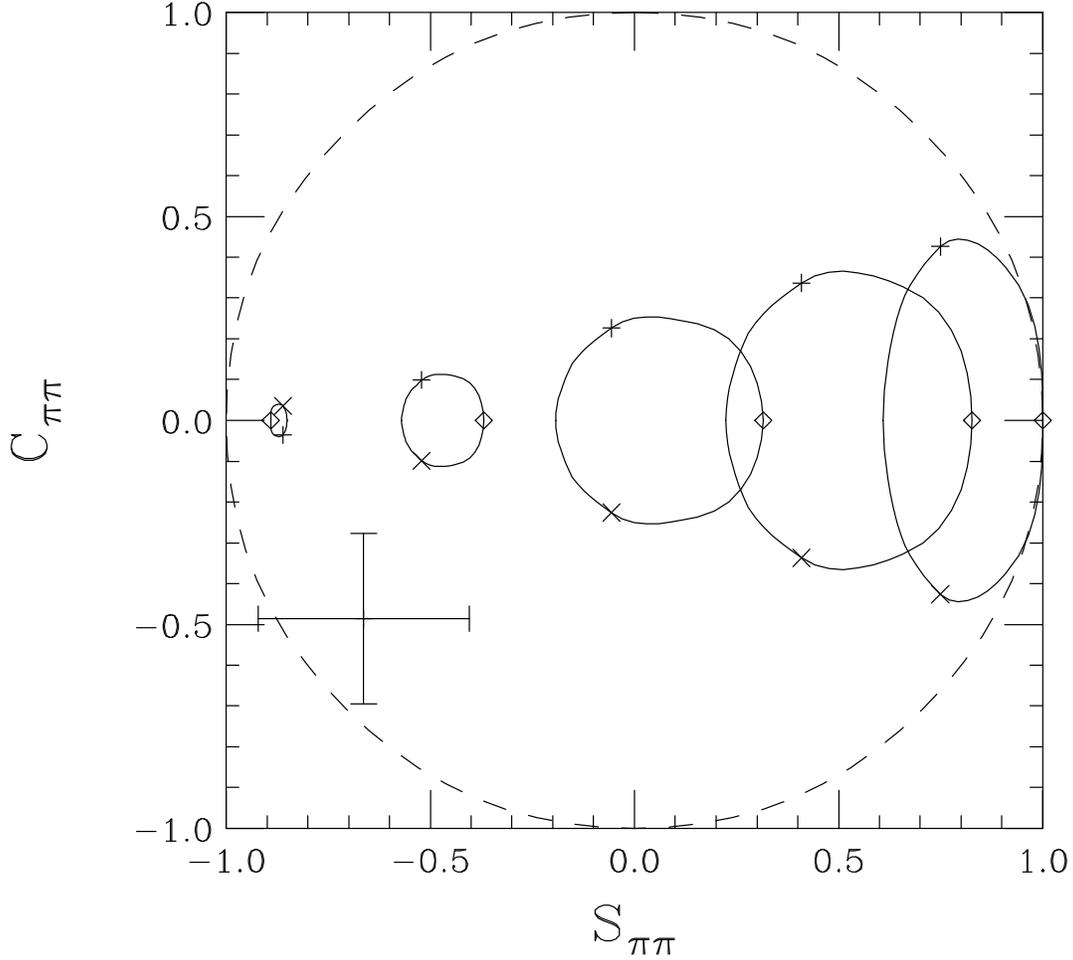}}
\caption{Values of $\spp$ and $\cpp$ for representative values of
$\alpha$ lying mostly outside the physical region.  Closed curves
correspond, from left to right, to $\alpha = -30^\circ$, $-15^\circ$,
$0^\circ$, $15^\circ$, and $30^\circ$.  Other notation as in
Fig.\ \ref{fig:cs00}. 
\label{fig:csu}}
\end{figure}

The corresponding plot for (mostly) unphysical values of $\alpha$ is
shown in Fig.\ \ref{fig:csu}.  If desired, one may map negative values of
$\alpha$ into the interval $[0,\pi]$ by the replacement $\alpha \to \alpha +
\pi$, $\delta \to \delta \pm \pi$, which leaves all expressions invariant.
The conventional physical region is bounded by $0 \le \alpha \le \pi -
\beta$.

The closed curves in Fig.\ \ref{fig:cs00} have considerable dependence on
$\delta$ for $\alpha$ around $\pi/2$.  One can show that $\spp$ becomes
independent of $\delta$
when $\cos 2 \alpha = |P/T|^2 \cos 2 \beta$.  Since $|P/T|^2$ is small,
these points are $\alpha \simeq \pi/4,~3 \pi/4$.  At such critical
values of $\alpha$ the curves degenerate into vertical lines.  
For $\alpha = \pi - \beta$, one has $\gamma = 0$, $\cpp = 0$,
$\spp = \sin 2 \alpha$, and the curves collapse to a point.

\begin{figure}
\centerline{\epsfysize = 5 in \epsffile{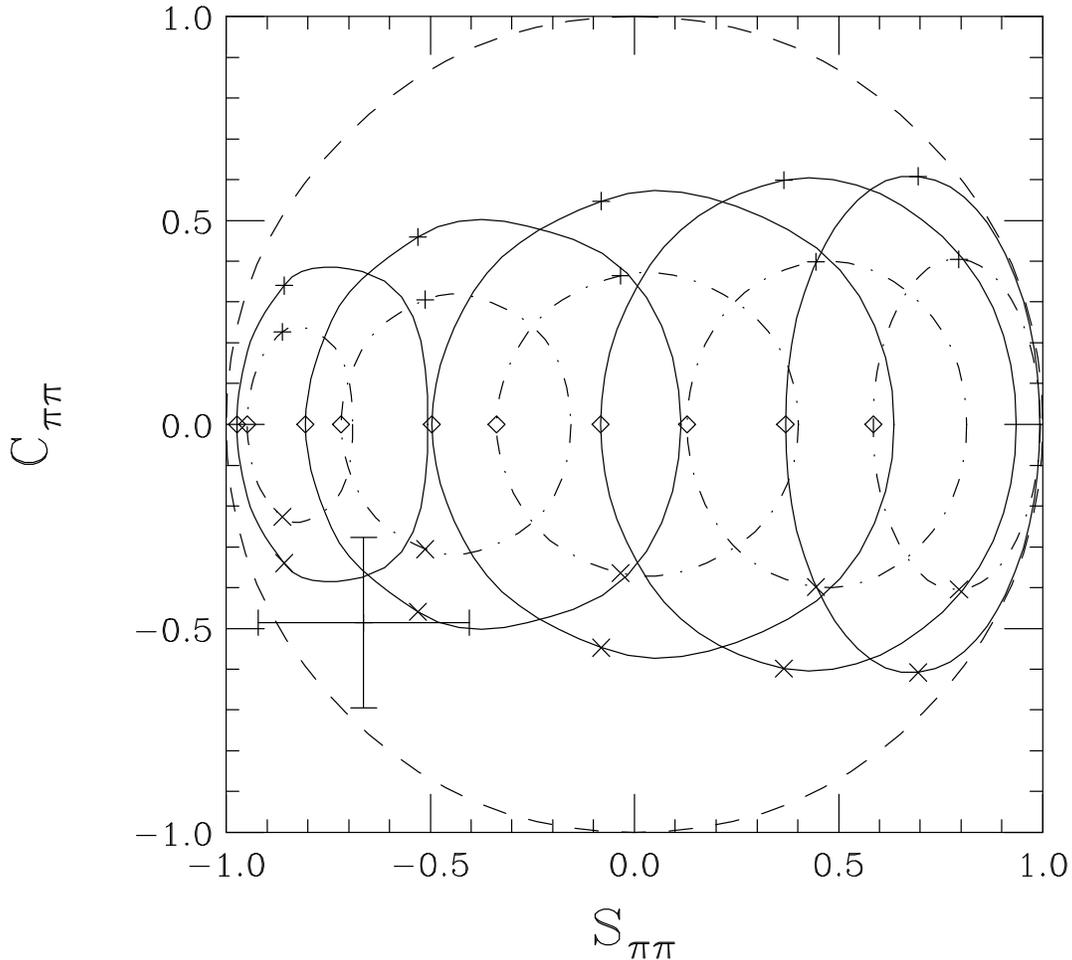}}
\caption{Values of $\spp$ and $\cpp$ as functions of $\alpha$ and
$\delta$; same as Fig.\ \ref{fig:cs00} except $|P/T| = 0.34$
(solid curves) and $|P/T| = 0.21$ (dot-dashed curves).
\label{fig:cs0+-}}
\end{figure}

The curves in Figs.\ \ref{fig:cs00} and \ref{fig:csu} were plotted for
the central values $\beta = 26^\circ$, $|P/T| = 0.28$.  Their dependence
on $\pm 1 \sigma$ variations of $\beta$ is quite mild for $\alpha$ in the
physical region,
while they are more sensitive to $\pm 1 \sigma$
excursions of $|P/T|$, as shown in Fig.\ \ref{fig:cs0+-}.

Let us imagine a measurement of $\spp$ and $\cpp$ which reduces present
errors by a factor of $\sqrt{3}$.  Given that the present measurements
are based on around a total of 100 fb$^{-1}$, one could envision such an
improvement when both BaBar and Belle report values based on 150 fb$^{-1}$.
Then the size of the error ellipse associated with $\spp$ and $\cpp$ will
be small in comparison with that of the closed curves for $\alpha$ in the
vicinity of $90^\circ$,
and measurement of these quantities could provide useful information
were it not for the fact that every point in the $\spp,\cpp$ plane
corresponds to several pairs $\alpha,\delta$.  The most important of
these pairs occurs when both values of $\alpha$ are in the physical
region but one corresponds to a certain value of $\delta$ and the other
(roughly) to $\pi - \delta$.  This discrete ambiguity is most severe
(corresponding to the most widely separated values of $\alpha$) when
$\cpp = 0$, corresponding to $\delta = 0$ or $\pi$.  For example, in
Fig.\ \ref{fig:cs00}, $\spp = \cpp = 0$ corresponds to both $\alpha
\simeq 76^\circ$ (when $\delta = 0$) and to $\alpha \simeq 105^\circ$ (when
$\delta = \pi$).  These values of $\alpha$ are separated by nearly
$30^\circ$.  We shall see in the next section how a measurement of the
branching ratio $\overline{{\cal B}}(B^0 \to \pi^+ \pi^-)$
can help resolve this ambiguity.

Measuring a nonzero value for $\cpp$ determines the sign of $\delta$,
but leaves an ambiguity between $\delta$ and $\pi - \delta$. The 
corresponding ambiguity in determining $\alpha$ becomes smaller when
$\delta$ moves away from 0 and $\pi$.
For maximal direct CP violation, corresponding to $|\delta| = \pi/2$,
one has $\sin \delta = \pm 1$, $\cos \delta = 0$, and no discrete
ambiguity.  These cases correspond to the envelope of the curves in
Figs.\ 1--3, joining the points labeled with $+$ ($\delta = \pi/2$)
or $\times$ ($\delta = - \pi/2$).

\section{Information from decay rate}

The quantity $\rpp$, defined in Eq.~(\ref{rpp}), can help resolve the 
discrete ambiguity
between $\delta = 0$ and $\delta = \pi$ in the case $\cpp = 0$, where such
an ambiguity is most serious.  It has been frequently noted \cite{sign}
that the central value of this quantity is less than 1, suggesting the
possibility of destructive interference between tree and penguin
amplitudes.  With the estimate $|T| = 2.7 \pm 0.6$ mentioned above, and
with the experimental average \cite{Iij} of CLEO, Belle, and BaBar
branching ratios equal to $\overline{{\cal B}}(B^0 \to \pi^+ \pi^-) =
(4.6 \pm 0.8) \times 10^{-6}$, we have $\rpp = 0.63 \pm 0.30$, which lies
suggestively but not conclusively below 1.  A value of 
$\rpp < 1$ would imply $\cos\delta < 0$ within the CKM framework,
since all currently allowed values of $\gamma$ correspond to $\cos\gamma > 
0$. Furthermore, a value of $\rpp$ below 1
permits one to set a bound on $\alpha + \beta$ or on $\gamma$, which is 
independent of $\delta$, 
\beq
\rpp = 1 + (|P/T| + \cos\delta\cos\gamma)^2 -  \cos^2\delta\cos^2\gamma
\geq \sin^2\gamma~,
\eeq
similar to the Fleischer-Mannel bound in $B\to K\pi$ \cite{FM}.
At the $1 \sigma$ level, this already implies $\gamma \le 71^\circ$ in the CKM
framework. In a more general framework, $\gamma \ge 109^\circ$ is also allowed.

\begin{figure}
\centerline{\epsfysize = 7 in \epsffile{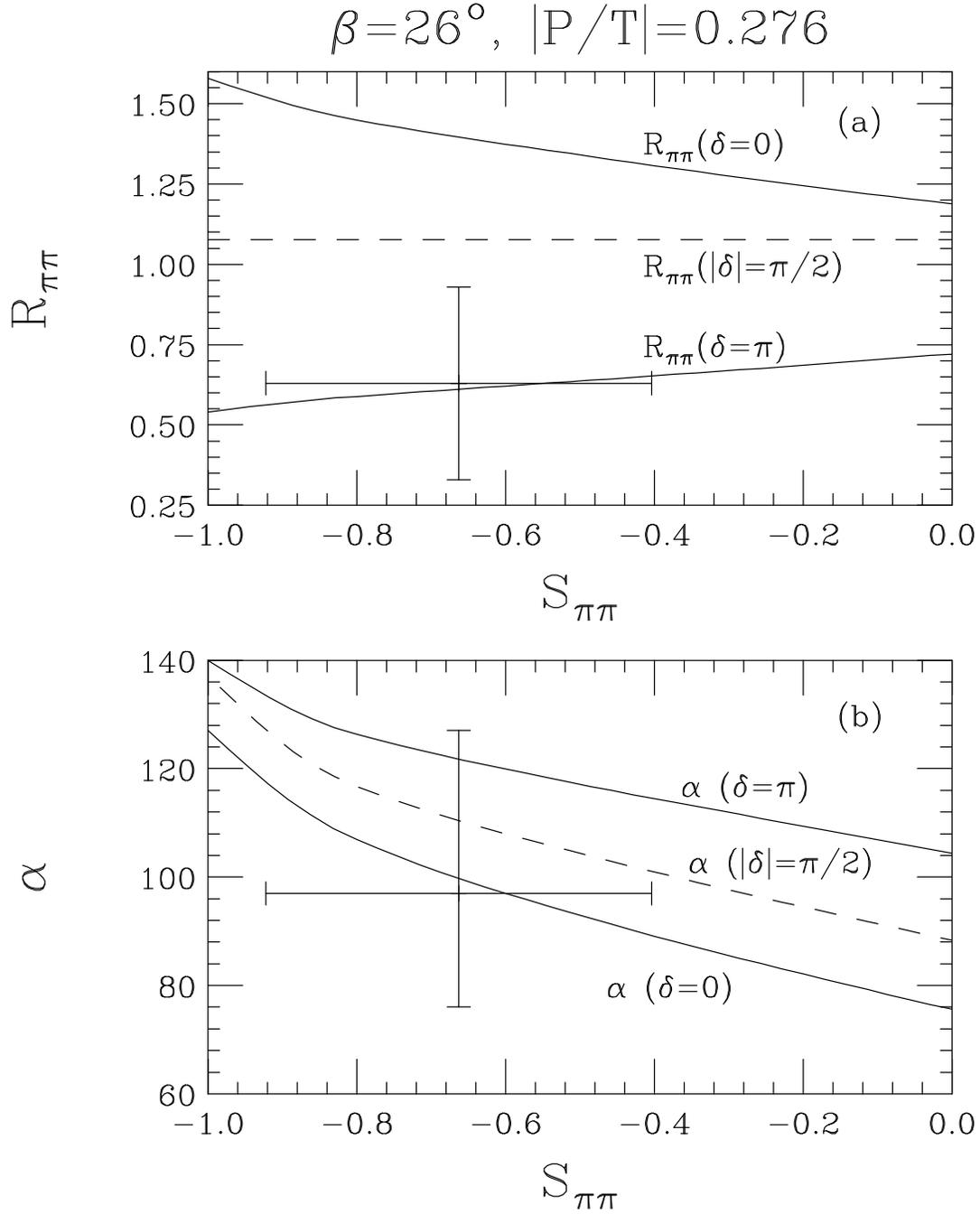}}
\caption{Values of (a) $\rpp$ and (b) $\alpha$ as functions of $\spp$ for the
cases $\delta = 0$ and $\delta = \pi$ leading to $\cpp = 0$ (solid
lines), and for $|\delta| = \pi/2$ (dashed lines).  The plotted points
correspond to experimental values of $\spp$ and (a) $\rpp$ or (b) $\alpha$.
Other parameters as in Fig.\ \ref{fig:cs00}.  For these sets of
parameters $\app < 0$; when $\cpp=0$ one has $\app = - (1 - \spp^2)^{1/2}$.
\label{fig:ra}}
\end{figure}

We show in Fig.\ \ref{fig:ra} the dependence of $\rpp$ and $\alpha$ on $\spp$
for the extreme cases $\delta = 0$ and $\delta = \pi$.  For reference we
also exhibit the curves for $|\delta| = \pi/2$.  As mentioned, only $\cpp$
depends on the sign of $\delta$.  Also shown are experimental points
corresponding to present ranges of $\rpp, \alpha$, and $\spp$.  If errors on
$\spp$ and $\cpp$ are reduced by about a factor of $\sqrt{3}$, and on $\rpp$
by a factor of about three,
as would be possible with a sample of 150 fb$^{-1}$ for each experiment,
one can see a constraint emerging which would favor one or the other
choices for $\delta$.  We discussed reduction of errors on $\spp$ and $\cpp$
already.  The corresponding reduction for $\rpp$ requires reduction of errors
on $|T|^2$ and $\overline{\cal B}(\bo \to \pi^+ \pi^-)$ from their present
values of 44\% and 17\%, respectively, each to about 10\%, which was shown
in Ref.\ \cite{LRpipi} to be possible with 100 fb$^{-1}$.

\section{Comparison with CKM parameters determined by other means}

\begin{figure}
\centerline{\epsfysize = 4.7 in \epsffile{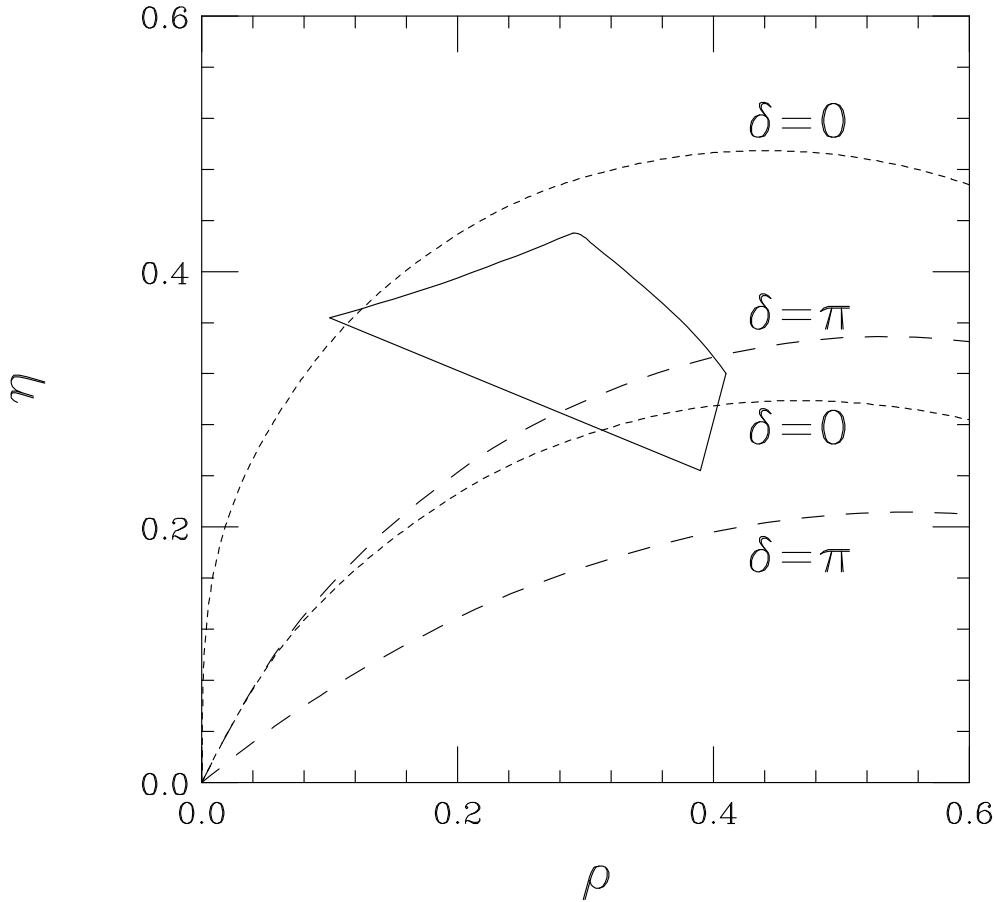}}
\caption{Plot in the $(\rho,\eta)$ plane of regions allowed by the
observed $1 \sigma$ ranges
$-0.92 \le \spp \le -0.40$ and $0.21 \le |P/T| \le 0.34$ for $\delta = 0$
(small dashes) and $\delta = \pi$ (large dashes), compared with region allowed
by other constraints (solid lines).  Bottom solid line:  lower bound on
$\beta$.  Upper left solid line:  upper bound on $\epsilon_K$.  Upper right
solid line:  upper bound on $|V_{ub}|$.  Right-hand solid line:  lower
bound on $\Delta m_d$.
\label{fig:rereg}}
\end{figure}

In Ref.\ \cite{GR01} we compared the region of CKM parameters allowed by
data on various weak transitions with that implied by the first observed range
of $\spp$ \cite{CSBa1} and $|P/T|$ for the case $\delta = 0$. 
In Fig.~\ref{fig:rereg} we reproduce that plot, corresponding to present
$1\sigma$ limits on $\spp$ and $|P/T|$ values in the range $0.21 \le |P/T| \le
0.34$,  along with the case $\delta = \pi$.

The case $\delta = \pi$ is seen to exclude a large region of the
otherwise-allowed parameter space, while $\delta = 0$ is compatible
with nearly the whole otherwise-allowed range.  Of course this does not
permit a distinction at present between the two solutions, but it
illustrates the potential of improved data.  Turning things around, the
examples in Fig.~\ref{fig:rereg} 
corresponding to $\delta =0$ and $\delta = \pi$ illustrate the importance 
of excluding one of these two values by means of the ratio $\rpp$.
Values $0 < |\delta| < \pi$ with $\cpp \ne 0$ correspond to constraints
intermediate between those for $\delta = 0$ and $\delta = \pi$.

The present $(\rho,\eta)$ constraints differ from those in Refs.\
\cite{GR01,LRpipi} based on the earlier BaBar data \cite{CSBa1}, which
were consistent (as are the present BaBar data \cite{CSBa2}) with vanishing
$\spp$ and $\cpp$.  In that case $\delta \simeq 0$ led to a significant
restriction in the $(\rho,\eta)$ plane, permitting only low values of $\rho$,
while $\delta \simeq \pi$ would have been consistent with nearly the whole
allowed $(\rho,\eta)$ region (as well as with the present data on $\rpp$).

\section{Information from width difference}

The quantity $\app$ appears with equal contributions in the time-dependent
decay rates of $B^0$ or $\ob$ to a CP-eigenstate, when the width difference
$\Delta\Gamma_d \equiv \Gamma_L - \Gamma_H$ between neutral $B$ mass
eigenstates is non-zero \cite{isi},

$$
\Gamma(B^0(t) \to \pi^+\pi^-) \propto e^{-\Gamma_d t}  
[ \cosh (\Delta\Gamma_d t/2) - \app\sinh (\Delta\Gamma_d t/2) 
\nonumber
$$
\beq
~~~~~~~~~~~~~~~~~~+ 
\cpp\cos(\Delta m_d t) - \spp\sin(\Delta m_d t) ]~.
\eeq
Width difference effects in the $B_s$--$\bar B_s$ system were investigated some 
time ago in time-dependent $B_s$ decays \cite{isi, isi2}.  The feasibility of
measuring corresponding $\Delta\Gamma_d$ effects in $B^0$ decays, expected 
to be much smaller but having a well-defined sign
$(\Delta \Gamma_d > 0)$ in the CKM framework, was
studied very recently \cite{Dighe}.  While a measurement of $\app$ in $B \to
\pi^+\pi^-$ is unfeasible in near-future experiments because of the very small
value of $\Delta\Gamma_d$ ($\Delta\Gamma_d/\Gamma_d < 1\%$), we will 
discuss the theoretical consequence of such a measurement.  This brief study
and its conclusion seem to be generic to a broad class of processes, including
the U-spin related decay $B_s(t) \to K^+K^-$ \cite{FL}, in which width
difference effects are much larger \cite{GR}.

In the absence of
$P$, one just has $\spp = \sin(2 \alpha)$, $\app = \cos(2 \alpha)$,
so the two quantities are out of phase with respect to one another
by $\pi/4$ in $\alpha$. 
This reduces part of the ambiguity in determining $\alpha$ from the 
mixing-induced asymmetry. The same is true when $\delta = 0$ or $\pi$,
since then $\alpha$ is replaced by $\alpha_{\rm eff}$ as noted in the
previous section.

The dependences of $\spp$ and $\app$ on $\delta$ for fixed $\alpha$ also
are out of phase with respect to one another, in the following sense.  When
$\spp$ is most sensitive to $\delta$, $\app$ is least sensitive, and
vice versa.  One can show, for example, that $\app$ is completely
independent of $\delta$ when
\beq
\sin 2 \alpha = - |P/T|^2 \sin 2 \beta~~~,
\eeq
which corresponds, since $|P/T|^2$ is small, to values of $\alpha$ near
0, $\pi/2$, and $\pi$.  Recall that the corresponding values for $\spp$ were
near $\pi/4$ and $3 \pi/4$.  Conversely, whereas $\spp$ is maximally
sensitive to $\delta$ near $\alpha = \pi/2$, $\app$ is maximally
sensitive to $\delta$ near $\alpha = \pi/4$ and $3 \pi/4$.

In the absence of the penguin amplitude $\app$ would just be $\cos 2
\alpha$. Since $\alpha$ is not too far from $\pi/2$ in its currently 
allowed range, $\app$ remains negative in this entire range
also in the presence of the penguin amplitude. Positive values of $\app$ are
obtained for values of $\alpha$ which are excluded in the CKM framework.
For the values $\delta = 0$ and $\delta = \pi$, when $\cpp = 0$, one 
has $\app = - (1 - \spp^2)^{1/2}$. For these values of $\delta$, $\spp$ is 
seen in Fig.~1 to lie in the range $-1.0 < \spp \le 1.0$, implying $-1.0 \le
\app \le 0$.  Since in the standard model one expects $\app$ to be 
negative, positive $\app$, obtained for
unphysical values of $\alpha$, would signify new physics. 

\section{Conclusions}

We have investigated the information about the weak phase $\alpha$ and the 
strong phase $\delta$ between penguin ($P$) and tree ($T$) amplitudes
which can be obtained from the quantities $\spp$ and $\cpp$ measured in the
time-dependent decays $\bo \to \pi^+ \pi^-$ and $\ob \to \pi^+ \pi^-$.  
One has a number of discrete ambiguities associated with the mapping
$(\spp,\cpp) \to (\alpha,\delta)$. These appear to be most severe when $\cpp
\simeq 0$, since very different values of $\alpha$ can be associated with
$\delta \simeq 0$ and $\delta \simeq \pi$. We have shown that under such
circumstances these ambiguities are resolved by sufficiently accurate
measurements of the ratio $\rpp$ of the flavor-averaged $\bo \to \pi^+ \pi^-$
branching ratio to its predicted value due to the tree amplitude alone.  At
present this ratio appears to be less than 1, but with large errors.  Reduction
of present errors on $\spp$ and $\cpp$ by a factor of $\sqrt{3}$ and on
$\rpp$ by a factor of three will have significant impact on these phase
determinations.  If a non-zero value of $\cpp$ is found, the discrete ambiguity
becomes less important, vanishing altogether when $|\delta| = \pi/2$.

A small value of $\rpp$, around its present central value, would favor
$\delta = \pi$ over $\delta = 0$, as shown in Fig.~\ref{fig:ra}(a). 
A large negative value of $\spp$, as indicated by the Belle measurement 
\cite{bCSBe}, favors large values of $\alpha$, in particular if $\delta \simeq
\pi$. This is demonstrated in Fig.~\ref{fig:cs0+-} and Fig.~\ref{fig:ra}(b).
Correspondingly, Fig.~\ref{fig:rereg} shows that low values of $\rho$ are 
excluded in the latter case. This figure, drawn also for the case $\delta =0$,
illustrates the important role of the measurement of $\spp$ and the
knowledge of $\delta$ in determining the CKM parameters $\rho$ and $\eta$.

Another parameter, called $\app$ here, equal to $\pm(1 - \spp^2
- \cpp^2)^{1/2}$, is measurable in principle in time-dependent
$\bo \to \pi^+ \pi^-$ decays if effects of the difference between
widths of mass eigenstates can be discerned.  The sign of $\app$
is enough to resolve a discrete ambiguity between values of $\alpha$
expected in the standard model (corresponding to $\app$ 
negative) and unphysical $\alpha$ (corresponding to $\app$ positive).

As has been noted previously \cite{sign}, there are hints of destructive
tree-penguin interference in $B^0 \to \pi^+ \pi^-$, which may be difficult to
reconcile with the favored range of CKM parameters without invoking large
values of $\delta$.  If this interesting situation persists, one may for the
first time encounter an inconsistency in the CKM description of CP violation,
which often assumes small strong phases.  Improved time-dependent measurements
of $B^0 \to \pi^+ \pi^-$ will be of great help in resolving this question.
Given that Standard Model fits \cite{Beneke,Hocker,Ciu} prefer $\cos \gamma>0$,
a value of $\rpp$ significantly less than 1 in the absence of any other
evidence for large $\delta$ also could call into question the applicability of
factorization to $B^0 \to \pi^+ \pi^-$ \cite{GR01}.  More accurate measurements
of the spectrum in $B \to \pi l \nu$ \cite{LRpipi} and more accurate tests of
factorization in other color-favored processes \cite{LRfact} will help to check
this possibility.

\section*{Acknowledgments}

We thank the CERN Theory Group and the organizers of the CERN Workshop on the
CKM Unitarity Triangle for hospitality during part of this work. The research
of J. L. R. was supported in part by the United States Department of Energy
through Grant No.\ DE FG02 90ER40560. This work was partially supported by
the Israel Science Foundation founded by the Israel Academy of Sciences and
Humanities and by the US - Israel Binational Science Foundation through
Grant No. 98-00237.

\def \ajp#1#2#3{Am.\ J. Phys.\ {\bf#1}, #2 (#3)}
\def \apny#1#2#3{Ann.\ Phys.\ (N.Y.) {\bf#1}, #2 (#3)}
\def \app#1#2#3{Acta Phys.\ Polonica {\bf#1}, #2 (#3)}
\def \arnps#1#2#3{Ann.\ Rev.\ Nucl.\ Part.\ Sci.\ {\bf#1}, #2 (#3)}
\def \art{and references therein}
\def \cmts#1#2#3{Comments on Nucl.\ Part.\ Phys.\ {\bf#1}, #2 (#3)}
\def \cn{Collaboration}
\def \cp89{{\it CP Violation,} edited by C. Jarlskog (World Scientific,
Singapore, 1989)}
\def \econf#1#2#3{Electronic Conference Proceedings {\bf#1}, #2 (#3)}
\def \efi{Enrico Fermi Institute Report No.}
\def \epjc#1#2#3{Eur.\ Phys.\ J.\ C {\bf#1}, #2 (#3)}
\def \f79{{\it Proceedings of the 1979 International Symposium on Lepton and
Photon Interactions at High Energies,} Fermilab, August 23-29, 1979, ed. by
T. B. W. Kirk and H. D. I. Abarbanel (Fermi National Accelerator Laboratory,
Batavia, IL, 1979}
\def \hb87{{\it Proceeding of the 1987 International Symposium on Lepton and
Photon Interactions at High Energies,} Hamburg, 1987, ed. by W. Bartel
and R. R\"uckl (Nucl.\ Phys.\ B, Proc.\ Suppl., vol. 3) (North-Holland,
Amsterdam, 1988)}
\def \ib{{\it ibid.}~}
\def \ibj#1#2#3{~{\bf#1}, #2 (#3)}
\def \ichep72{{\it Proceedings of the XVI International Conference on High
Energy Physics}, Chicago and Batavia, Illinois, Sept. 6 -- 13, 1972,
edited by J. D. Jackson, A. Roberts, and R. Donaldson (Fermilab, Batavia,
IL, 1972)}
\def \ijmpa#1#2#3{Int.\ J.\ Mod.\ Phys.\ A {\bf#1}, #2 (#3)}
\def \ite{{\it et al.}}
\def \jhep#1#2#3{JHEP {\bf#1}, #2 (#3)}
\def \jpb#1#2#3{J.\ Phys.\ B {\bf#1}, #2 (#3)}
\def \lg{{\it Proceedings of the XIXth International Symposium on
Lepton and Photon Interactions,} Stanford, California, August 9--14, 1999,
edited by J. Jaros and M. Peskin (World Scientific, Singapore, 2000)}
\def \lkl87{{\it Selected Topics in Electroweak Interactions} (Proceedings of
the Second Lake Louise Institute on New Frontiers in Particle Physics, 15 --
21 February, 1987), edited by J. M. Cameron \ite~(World Scientific, Singapore,
1987)}
\def \kaon{{\it Kaon Physics}, edited by J. L. Rosner and B. Winstein,
University of Chicago Press, 2001}
\def \kdvs#1#2#3{{Kong.\ Danske Vid.\ Selsk., Matt-fys.\ Medd.} {\bf #1}, No.\
#2 (#3)}
\def \ky{{\it Proceedings of the International Symposium on Lepton and
Photon Interactions at High Energy,} Kyoto, Aug.~19-24, 1985, edited by M.
Konuma and K. Takahashi (Kyoto Univ., Kyoto, 1985)}
\def \mpla#1#2#3{Mod.\ Phys.\ Lett.\ A {\bf#1}, #2 (#3)}
\def \nat#1#2#3{Nature {\bf#1}, #2 (#3)}
\def \nc#1#2#3{Nuovo Cim.\ {\bf#1}, #2 (#3)}
\def \nima#1#2#3{Nucl.\ Instr.\ Meth.\ A {\bf#1}, #2 (#3)}
\def \np#1#2#3{Nucl.\ Phys.\ {\bf#1}, #2 (#3)}
\def \npps#1#2#3{Nucl.\ Phys.\ Proc.\ Suppl.\ {\bf#1}, #2 (#3)}
\def \os{XXX International Conference on High Energy Physics, Osaka, Japan,
July 27 -- August 2, 2000}
\def \PDG{Particle Data Group, D. E. Groom \ite, \epjc{15}{1}{2000}}
\def \pisma#1#2#3#4{Pis'ma Zh.\ Eksp.\ Teor.\ Fiz.\ {\bf#1}, #2 (#3) [JETP
Lett.\ {\bf#1}, #4 (#3)]}
\def \pl#1#2#3{Phys.\ Lett.\ {\bf#1}, #2 (#3)}
\def \pla#1#2#3{Phys.\ Lett.\ A {\bf#1}, #2 (#3)}
\def \plb#1#2#3{Phys.\ Lett.\ B {\bf#1}, #2 (#3)}
\def \pr#1#2#3{Phys.\ Rev.\ {\bf#1}, #2 (#3)}
\def \prc#1#2#3{Phys.\ Rev.\ C {\bf#1}, #2 (#3)}
\def \prd#1#2#3{Phys.\ Rev.\ D {\bf#1}, #2 (#3)}
\def \prl#1#2#3{Phys.\ Rev.\ Lett.\ {\bf#1}, #2 (#3)}
\def \prp#1#2#3{Phys.\ Rep.\ {\bf#1}, #2 (#3)}
\def \ptp#1#2#3{Prog.\ Theor.\ Phys.\ {\bf#1}, #2 (#3)}
\def \rmp#1#2#3{Rev.\ Mod.\ Phys.\ {\bf#1}, #2 (#3)}
\def \rp#1{~~~~~\ldots\ldots{\rm rp~}{#1}~~~~~}
\def \si90{25th International Conference on High Energy Physics, Singapore,
Aug. 2-8, 1990}
\def \slc87{{\it Proceedings of the Salt Lake City Meeting} (Division of
Particles and Fields, American Physical Society, Salt Lake City, Utah, 1987),
ed. by C. DeTar and J. S. Ball (World Scientific, Singapore, 1987)}
\def \slac89{{\it Proceedings of the XIVth International Symposium on
Lepton and Photon Interactions,} Stanford, California, 1989, edited by M.
Riordan (World Scientific, Singapore, 1990)}
\def \smass82{{\it Proceedings of the 1982 DPF Summer Study on Elementary
Particle Physics and Future Facilities}, Snowmass, Colorado, edited by R.
Donaldson, R. Gustafson, and F. Paige (World Scientific, Singapore, 1982)}
\def \smass90{{\it Research Directions for the Decade} (Proceedings of the
1990 Summer Study on High Energy Physics, June 25--July 13, Snowmass,
Colorado),
edited by E. L. Berger (World Scientific, Singapore, 1992)}
\def \tasi{{\it Testing the Standard Model} (Proceedings of the 1990
Theoretical Advanced Study Institute in Elementary Particle Physics, Boulder,
Colorado, 3--27 June, 1990), edited by M. Cveti\v{c} and P. Langacker
(World Scientific, Singapore, 1991)}
\def \TASI{{\it TASI-2000:  Flavor Physics for the Millennium}, edited by J. L.
Rosner (World Scientific, 2001)}
\def \yaf#1#2#3#4{Yad.\ Fiz.\ {\bf#1}, #2 (#3) [Sov.\ J.\ Nucl.\ Phys.\
{\bf #1}, #4 (#3)]}
\def \zhetf#1#2#3#4#5#6{Zh.\ Eksp.\ Teor.\ Fiz.\ {\bf #1}, #2 (#3) [Sov.\
Phys.\ - JETP {\bf #4}, #5 (#6)]}
\def \zpc#1#2#3{Zeit.\ Phys.\ C {\bf#1}, #2 (#3)}
\def \zpd#1#2#3{Zeit.\ Phys.\ D {\bf#1}, #2 (#3)}

\end{document}